\documentstyle[twoside,fleqn,espcrc2]{article}
\def\undertext#1{\vtop{\hbox{#1}\kern 1pt \hrule}}

\def\tr{\hbox{tr}\,}

\def\be{\begin{equation}}
\def\ee{\end{equation}}
\def\bea{\begin{eqnarray}}
\def\eea{\end{eqnarray}}
\def\eqref#1{(\ref{#1})}
\newcommand{\AmS}{{\protect\the\textfont2
  A\kern-.1667em\lower.5ex\hbox{M}\kern-.125emS}}

\begin{document}

\title{
\begin{flushright}
CERN-TH-6754/92  \\
December 1992
\end{flushright}
D-dimensional Induced Gauge Theory as a Solvable
       Matrix Model}

\author{V.A.Kazakov
\address{Theory Division of CERN,
        CH-1211 Geneva 23, Switzerland \\
        and  \\
                 Laboratoire de Physique Th\'eorique   \\
        de l'Ecole Normale Sup\'erieure, 24 rue Lhomond,
            75231 Paris Cedex 05, France         }}

\begin{abstract}
We discuss basic features and new developments in   recently proposed
induced gauge theory \cite{KM}  solvable in any number of dimensions in
the
limit of infinite number of colours.  Its geometrical (string) picture
is clarified, using planar graph expansion of the corresponding matrix
model. New  analytical approach is proposed for this theory which is
based on its equivalence to an effective two-matrix model. It is
shown on some particular examples
how the approach works.
               This approach may be applicable
to a wide class of matrix models with tree-like quadratic couplings
     of matrices.

 (This talk was presented on the International Symposium on Lattice
 Field Theory "LATTICE-92" in Amsterdam, the Netherlands, 15-19
 September 1992)

\end{abstract}

\maketitle

\section{Introduction}

The recently proposed D-dimensional
induced gauge theory solvable in the
large N limit \cite{KM} has virtually produced some ambitious
hopes
to provide the long searched Master Field (MF) solution for QCD. This
means      that we would find a   set of field variables that  become
classical in this limit. In principle, we know such a set: Wilson loop
functionals, which satisfy classical loop equations (Makeenko-Migdal
equations). But owing to their complicated functional character and
essential non-locality they do not advance us very  much on the way to an
analytical solution of multicolour QCD.

      Another idea which so far was  developed  separately in field
theory, was the so called induced gauge theory.  Brifly the idea of
induction is the following: let us start from a Lagrangian of a field
theory containing only   massive
gauged matter of some kind, with no addition of
any kinetic terms for the corresponding gauge field. If we integrate
out the matter we obtain a (generally) non-local effective action for
only gauge fields. If one then increases the mass of matter field one
obtains, in principle, the 1/(mass) expansion of the effective action
in terms of local operators.

     Of course, things are not as simple as they seem at    first
sight. We have to satisfy a few conditions to induce  a needed gauge
theory: the sign of the coupling of an induced interaction should be
correct, the power-like divergences should be absent (to be left with
only a logarithmic dependence of the induced gauge coupling on the
cut-off), non-perturbative (e.g.  in all orders of 1/(mass) expansion)
locality of the induced action, e.c.

   Surprisingly enough, one can formulate a gauge theory of this type:
on the one hand, this induces (at least naively, in the logarithmic
approximation) the Yang-Mills gauge coupling, and on the other hand,
it is solvable by means of the large N matrix model technique. The
MF appears to be local in this case, which immediately raises   some
doubts on the possibility for this theory to serve as  a right QCD
realization in 4 dimensions. And indeed, as    will be clear from the
following, the theory obeys symmetry too much   (with respect to
the centre of the gauge group) to describe a real QCD
 \cite{Sem},         \cite{BouSym}.  It is doubtful
     that this symmetry can be broken spontaneously, unless we make
the model more complicated (and thus unsolvable).

   Nevertheless, it seems to be interesting in itself
    that a solvable non-abelian gauge theory
of any kind exists in physical dimension.

 First of all, the search for an MF for real QCD along these
     lines should not be necessarily a hopeless enterprise. Some
attempts of this kind are on the way \cite{MaKh,MigFerm}.

 Secondly,
the corresponding D-dimensional matrix model might be a meaningful
string theory with an infinite mass spectrum of physical states.
   In particular,                  the analytical methods proposed
for the investigation of this model can be used for the solution of
any multi-matrix model with the tree-like quadratic form of couplings
between the matrices (which includes, say, Q-component Potts models on
random graphs) \cite{Kpot}.       Moreover,       the induced gauge
model under consideration is itself the model embedded into a tree,
rather than into a real D-dimensional space, at least in the large-N
limit \cite{Bou}. For large N the space almost falls out from the
model and its dimensionality appears only through the number of
nearest neighbours
       of the D-dimensional lattice. In the next orders of 1/N, the
space restores little by little, but the trivial space structure of
the lowest order also signals that it is not a QCD realization.

On the other hand, as a matrix model,             this theory obeys
all the features of  some string theory, which means that it is a
theory of an extended object from the point of view of the internal
(transversal) modes of the world sheet given by the corresponding planar
graphs. Therefore, amid the tree nature of the model in the embedding
space, it can produce an infinite and appropriately scaled spectrum
of physical states due to these transversal modes. So, it is worth
trying it.

An         analytic   approach   of the model  was  suggested in
\cite{KM}, and a considerable progress was achieved in the papers
\cite{Mig},              where        the critical behaviour of the
model was established. In the papers \cite{Gross}  (see also
\cite{MakLoop,Casel}), the
case of Gaussian "induction matter" was solved exactly and the 1D
string theory solution was demonstrated.  In the paper \cite{Bou}
the new critical behaviour of this model was claimed, with a
log-of-log singularity of the string susceptibility. This result
contradicts in a way    the results of \cite{Mig},  so that the whole
question of critical behaviour in the model needs to be clarified,
in our  opinion.

    In sect.2 we will formulate the model in the continuum and on the
lattice, describe the induction idea and discuss the planar graphs
picture.

In section 3,
 the issue of the additional symmetry of the centre of the
gauge group is reviewed.

   In sect.4   a general master field equation (MFE) will be formulated
and the model will be shown to be equivalent to a random surface
embedded in a tree in the large-N approximation.

   In sect.5 the quantitative approach to the model will be formulated,
based on its  equivalence to the 2-matrix model (2MM), with the
self-consistent effective potential for each of matrices. Using the
orthogonal polynomials approach one can then write down a simple
functional
(and, in principle, solvable) equation.
   As an application the induced gauge    model with a Gaussian scalar
potential for any dimension
will be solved in section 6 by this method and the result  will coinside
with the known one.

Sect.7 will review existing results and proposals on the induced gauge
theory.

   \section{Formulation of the model and
   its physical interpretation}

  In the continuum version the model under consideration is just a
gauged scalar field theory in the
adjoint representation of the group SU(N)
with the Lagrangian:
\begin{equation}
L= \frac{N}{g_0^2} \tr  \left(\left(\partial_{\mu} \Phi +
i \left(A_{\mu},\Phi\right)\right)^2 +   V(\Phi) \right)
\label{Contin}
\end{equation}
Here $\Phi(x)$ is an NxN matrix Hermitian field (a scalar field in
the adjoint representation of SU(N)), and $A_\mu(x)$ is a gauge field.

Note that one does not add the Yang-Mills interaction explicitly.

   The action for the lattice version of the same model looks as:
\begin{equation} \matrix{
S =  \sum_x N \, \tr      [ V(    \Phi  (x))    \cr
- \sum_{\mu= 1, 2,...D}
     (\Phi(x)-U_{\mu}(x)\Phi(x+\mu) U_{\mu}^{\dagger}(x))]
}
\label{Action}
\end{equation}

The partition function is given by:

\begin{equation}
\int \prod_x d^{N^2}\Phi_x \int \prod_{<xy>} (dU)_{SU(N)} \, \exp{-N S}
\label{Partlat}
\end{equation}

 The scalar potential can be an arbitrary polynomial
\begin{equation}
V(\Phi) = m^2 \Phi^2 + \lambda_0 \Phi^4 + ...
\label{poten}
\end{equation}
or even a more complicated function.

Note that no pure Wilson plaquette term was added in (\ref{Action}).

Formally this model can induce the Yang-Mills interaction, if one would
believe that the parameters of the scalar potential $V(\Phi)$ could
be adjusted so as to get into the regime of asymptotic freedom. Of course
This is not guaranteed at all                         within this model.
Considering this possibility we note that already the simplest one-loop
logarithmic contribution induces the Yang-Mills interaction:
\begin{equation}
\int DA D\Phi \exp(-S) \propto \int DA \exp\left(-S_{ind}[U]\right)
\label{Induced}
\end{equation}
where the first terms of the
$1/m$ expansion ($m^2=m_0^2-m^2_{crit}$) look as
\begin{equation}
L_{ind}(A) = \frac{1}{g^2_{ind}}\tr F_{\mu\nu}^2 +\frac{const}{m^2}
\tr(\partial_\mu F_{\mu\nu})^2 + ...
\label{Indac}
\end{equation}
with the induced gauge coupling given by
\begin{equation}
\frac{1}{ g^2_{ind}}
    =      \frac{N}{96 \pi^2}\ln \frac{\Lambda^2}{m^2}
\label{ScalarLoop}
\end{equation}
{}From the last formula, we see that the renormalized mass $m^2$ must be
chosen in such a way that it would be much smaller than the original
               cut-off scale $\Lambda$,  and much larger than
the physical mass scale        $\mu$        of (would be) glueballs.
The latter is given by the asymptotic freedom relation:
\begin{equation}
\ln \frac{m^2}{\mu^2}  \rightarrow  \frac{48 \pi^2}{11Ng_0^2}
\label{AsymptoticFreedom}
\end{equation}
Comparing these two relations we find the scaling law:
\begin{equation}
\mu^2 \rightarrow  ( m_0^2 - m_c^2)^b
\label{ScalingLaw}
\end{equation}
where
\begin{equation}
b = {\frac{23}{22}}
\label{b}
\end{equation}
One should not take              this number too seriously.
        This naive scaling law was obtained without taking into account
the      corrections from the scalar field self-interactions and from the
effects of hard gluons (note that higher derivative terms in the
induced action  (\ref{Indac}) are suppressed by $1/m^2$, and not by
$1/\Lambda$, and hence will contribute to the renormalization of the
scaling exponent).

The correct calculation of this scaling law, as well as the whole issue
of the existence of the correct physical QCD phase in our theory is
so far beyond our technical possibilities. May be, computer simulations,
such as those started in \cite{MonteC},       can shed some light on it.
In the  opinion of the author, it is unlikely   that this phase could
correspond to any local  large-N master field approach described here,
but the possibility for this phase to exist for finite N cannot be
completely excluded.

\section{ Planar graph representation and extra $Z_N$ symmetry}
Let us try to give some geometrical interpretation of our theory in
terms of planar graph expansion, since we will try later to give
   it the meaning of some improved bosonic string, and the planar
graphs usually play the role of a regularization of a world sheet
of a string theory.

Let us integrate first over the scalar field        in (\ref{Contin}).
In the large N limit we will get a standard planar Feynman diagram
technique, as for scalar field theory, with $\Phi^3,\Phi^4...$-vertices,
but the propagators  $G^{ij}_{kl}(x,y,A)$
will be modified owing to the external gauge field:
they will be given by sums over paths of P-ordered Wilson factors in
adjoint representation:
\begin{equation}  \matrix{
                      G^{ij}_{kl}(x,y,A)  =  \cr
\int D\Gamma(x,y) \exp{-m^2 Length(\Gamma)}    \cr
\left(P\exp[i     \int_{\Gamma(x,y)}dz_\mu A_\mu(z)] \right)^{ij} \cr
\left(P\exp[ -i   \int_{\Gamma(x,y)}dz_\mu A_\mu(z)] \right)_{kl}
}
\label{Propagator}
\end{equation}
    The whole planar graph expansion can be represented in this way
as standard planar graphs for a scalar field "embedded" into the
"external  gauge field media".

The integration that is left, over the gauge field, would produce some
non-local interactions of different pieces of this discretized world
sheet. In principle, we could try to integrate over them in the
lattice version of the same representation, using the technique
proposed in \cite{KaQCD,KoQCD},  in order to get some final
representation of this model in terms of random surfaces, but
it would give any promising calculational ideas.

Let us note only  that on the lattice one can immediately see how
the term like the Wilson plaquette action arises in the strong coupling
expansion over the kinetic term in the induced gauge theory. Four
terms corresponding to the links around    a plaquette give after
a   Gaussian scalar field integration:
\begin{equation}
   |\tr U(plaquette)|^2 \rightarrow
    N^2 + \frac{1}{2} a^4 \tr F_{\mu\nu}^2
\label{Wilson}
\end{equation}

As a consequence of the adjoint representation used in our model, we
will always get the induced action for the gauge field with every
matrix element of
gauge $U$-matrix multiplied by  some matrix element of
 its conjugate $U^+$. As a result we
have          an extra gauge $Z^N$ symmetry in our model, the symmetry
with respect to
   the centre of the       $SU(N)$ gauge group. Namely, one can rotate
every $U$ matrix by the Abelian $Z_N$ factor:
\begin{equation}
U_{xy} \rightarrow U_{xy} \exp{i\frac{2\pi}{N} k_{xy}}
\label{Center}
\end{equation}
where $k_{xy}$ is an integer, with
the consequnce                        that any Wilson average $W(C)$
in the fundamental representation will be
non-zero (and equal to 1) iff the loop $C$ forms a tree in the x-space
(with zero minimal area of the surfaces spanned on it):
\begin{equation}
W(C) = <1/N \tr U(C)> = \delta_{0,Area(C)}
\label{Area}
\end{equation}
The same will be true for the adjoint Wilson loop in the large N limit,
since it factorizes in this limit into two fundamental loop averages.

Of course, this kind of superconfinement has nothing in common with
the physical confinement and expected area law for fundamental loops.

         A way out                could be some mechanism of
  spontaneous breaking of this symmetry. Even though this is a gauge
symmetry, one may hope that it may be broken at least in the large N
limit. To analyse this possibility  we can use the arguments of
\cite{MaKh}.  For example, for one plaquette action in an adjoint
representation we can use the large N factorization property:
\begin{equation} \matrix{
S_{adj}  = \cr
 \sum_{plaquettes}    |U(plaquette)|^2     \cr
=_{N  \rightarrow \infty}
N\beta <\frac{1}{N} \tr U(plaquette)>       \cr
\sum_{plaquettes}  \tr U(plaquette)
}
\label{Factoract}
\end{equation}
which gives a non-linear equation for the effective coupling $\bar{\beta}$
\begin{equation}
\bar{\beta} = \beta W(plaquette,\bar{\beta})
\label{Nonlin}
\end{equation}
 The
  trivial solution
\begin{equation}
\bar{\beta} = 0 ,       W(plaquette,\bar{\beta}) =0
\label{Triv}
\end{equation}
corresponds to the non-broken $Z_N$ symmetry, where as a   non-trivial
solution
\begin{equation}
\bar{\beta} \neq   0,   W(plaquette,\bar{\beta}) \neq   0
\label{Nontriv}
\end{equation}
corresponds to the    broken $Z_N$ symmetry with a possibility of
the correct physical behaviour.

It is easy to generalize this picture to the whole induced action
of our model, which will include the sums over all loops, and not
only over plaquettes.

One should stress that the corresponding large N phase transition
with this symmetry breaking should take place before we approach the
continuous limit of the theory.

This scenario seems to be unlikely   for the large N master field
approach presented below. It seems that this centre of the group
symmetry remains unbroken      for infinite N.

\section{Master field equation for the eigenvalues of scalar field}

In order to see that our model is exactly solvable in the large N limit
let us     choose an opposite order of integration in the functional
integral: first we will  integrate over gauge fields, and then over
the (eigenvalues of the) scalar field.

First we demonstrate the idea in the continuum version, but a rigorous
treatment will be possible only on the lattice.

We introduce, as usual, the "angular"  parametrization of scalar field
in terms of eigenvalues $\phi=diag(\phi_1,\phi_2,...,\phi_N)$ and
eigenfunctions $\Omega_{ij} \epsilon SU(N)$:
\begin{equation}
\Phi(x)=\Omega^+(x)\phi(x)  \Omega(x)
\label{Angular}
\end{equation}
Putting it into the Lagrangian (\ref{Contin}) we obtain:
\begin{equation} \matrix{
L=       N \sum_{k=1}^{N}
       \left((\partial_{\mu} \phi_k)^2  + V(\phi_k) \right)  \cr
+ \sum_{i,j=1}^N (\phi_i - \phi_j)^2     |B_\mu^{ij}|^2
}
\label{Gaugerot}
\end{equation}
where
\begin{equation}
B_\mu =         \Omega^+ A_\mu      \Omega(x)
    +  i\Omega^+\partial_\mu \Omega(x)
\label{Gauchoise}
\end{equation}
The field $B_\mu$ is just a gauge rotated $A_\mu$, therefore, naively
speaking, the integral over $B_\mu$ is Gaussian. Integrating over it
we would obtain a simple effective action for only the eigenvalues,
ready for the application of the large N master field approach. But the
situation is slightly more complicated: any change of variables:
$\Omega \rightarrow P \Omega$, $\phi \rightarrow P^+ \phi P$, where $P$
is a permutation matrix, does not change $\Phi=\Omega^+\phi\Omega$.
Therefore, integrating independently over $B_\mu$, we would overcount
the possible configurations of scalar fields.

To avoid overcounting we can, say, impose the condition $\phi_1<
\phi_2<...<\phi_N$, or we have to  subtract one by one   the overcounted
configurations. The second possibility can be naturally realized in
the lattice version.

Returning to the lattice  let us note, that after a   change of variables
(\ref{Angular}) at every vertex of the hypercubic lattice, we can again
choose the gauge in such a way that the angular degrees of freedom
will  be absorbed into the gauge fields: $\Omega_x U_{xy} \Omega^+_x
\rightarrow U_{xy}$, because of  the group invariance of the Haar measure
(on every link of the lattice separately). Then the integral over
every link gauge variable can be performed by means of the so-called
Itzykson-Zuber-Kharish-Chandra formula \cite{IZ}:
\begin{equation}  \matrix{
I(\phi,\chi) =
\int DU  \exp \left(N    \tr(\phi-U \chi U^{\dagger})^2\right) \cr
\propto \frac{\det_{ij}\exp(N(\phi_i-\chi_j)^2)}{\Delta(\phi)\Delta(\chi)}
}
\label{IZ}
\end{equation}
where
\begin{equation}
\Delta(\phi) = \prod_{i<j} (\phi_i-\phi_j)
\label{Vandermond}
\end{equation}
Taking into account extra $\Delta^2(\phi(x))$ of the Dyson measure for
the "angular" parametrization  of the scalar field, we arrive at the
partition function:
\begin{equation}
Z=\int \prod_{x,k} d\phi^{(k)}_x\exp( - S_{eff})
\label{EffPar}
\end{equation}
where
\begin{equation}  \matrix{
S_{eff} = \sum_{xy} \log I(\phi_x,\phi_y) +   \cr
\sum_x \left(\log \Delta^2(\phi_x) -N \sum_{k=1}^N V(\phi_x^{k})\right)
}
\label{EffAc}
\end{equation}
The above mentioned naive continuum action (\ref{Gaugerot}) after
    integration over $B_\mu$, will correspond only to the diagonal
term in the determinant in (\ref{IZ}). All other $(N!-1)$ terms can be
considered as a consequent subtraction of the overcounted permutations
of eigenvalues. It would be interesting to find some continuous
field theoretical description of this object as a generalization of
an ordinary spatial derivative.

Let us remark that in spite the Van-der-Monde determinants in the
denominator of (\ref{IZ}), the action (\ref{EffAc}) is not singular
at all with respect to the coinciding eigenvalues: the determinant
in the numerator has zeros  cancelling these singularities, since
the whole integral evidently has no such singularities.

Now our effective action (\ref{EffAc}) depends only on N eigenvalues
and we are in position to apply the saddle-point method in the large
N limit and derive the corresponding saddle-point equation.
Namely, we look for the classical configuration for the eigenvalues
obeying the stationarity condition (MFE):
\begin{equation}
\frac{\partial  S_{eff} }{\partial \phi_x^i} = 0
\label{Stationary}
\end{equation}
{}From the physical point of view it is natural to expect the spatially
homogeneous vacuum for this system, so, identifying

\begin{equation}
\phi^i_x = \phi_i = const(x)
\label{Homogen}
\end{equation}
we can write the effective action as
\begin{equation}       \matrix{
S_{eff} = (volume)(2D)     [(\log I(\phi  ,\phi  ) +    \cr
\frac{1}{2D}       \log \Delta^2(\phi  )    \cr
      - N \frac{1}{2D}\sum_{k=1}^N V(\phi_k)  ]
}
\label{HomAc}
\end{equation}
Now the                        MFE  (\ref{Stationary}) reads as
\begin{equation}
2D \frac{\partial \log I(\phi,\phi)}{\partial \phi_k} +
2 \sum_i '\frac{1}{\phi_k - \phi_i} =  V'(\phi_k)
\label{MFESum}
\end{equation}
or, in terms of     density of     eigenvalues,
\begin{equation}
\rho(\phi) =\frac{1}{N}\frac{dk(\phi)}{d\phi}
\label{Density}
\end{equation}
the basic quantity to be calculated in this approximation:
\begin{equation}
2D \frac{\partial \log I(\phi,\phi)}{\partial \phi  } +
2 P\int d\mu \rho(\mu) \frac{1}{\phi   - \mu   } =  V'(\phi  )
\label{MFEInt}
\end{equation}
But we have not yet closed the system of equations, since we still have
to find some effective approach for the calculation of $I(\phi,\phi)$.
The original formula of Itzykson and Zuber (\ref{IZ}) is valid for
any N, but it is not very suitable in the limit of large N, since we
have to deal with the sum of N! sign-changing terms.

An         approach   to this problem was sketched in \cite{Mig},
where the integral equation for the calculation of the IZ-integral in
the large N limit was derived. Later we will discuss this approach.

In the next section we will propose a new approach based on the
effective                 2MM.

To conclude this section, let us recall a comment, made in \cite{Bou},
on the equivalence of our induced gauge model and a multi-matrix model
embedded into an infinite Bethe tree      with the coordination number
2D, in the large N limit. This Q-matrix model obeys the following action:
\begin{equation}
S=N \tr\left(\sum_{<ij>}     \Phi_i \Phi_j + \sum_{i=1}^QV(\Phi_i)\right)
\label{Tree}
\end{equation}
where the first sum runs over the bonds of an infinite tree with the
coordination number 2D, and the second   over its vertices. Note that
if we would gauge the model in the spirit of our induced gauge theory,
it would not change it since, on any tree, the gauge variables can be
completely absorbed into the angular degrees of freedom of matrices.

On the other hand, integrating by means of the IZ-integral over the
relative "angles" on every bond, and looking for the homogeneous
saddle point for the eigenvalues (the infinite Bethe tree is certainly
translational  invariant), we obtain the same effective action
(\ref{HomAc}) and the same MFE (\ref{MFESum}) or (\ref{MFEInt})
as for the D-dimensional induced gauge model.

 Hence, in the large N limit the models coincide. Of course this
 coincidence is very worrysome from the point of view of a possible
       equivalence with QCD. As we see the dependence on the space
structure almost disappears from the problem in the main order in 1/N.
Its dimensionality D enters the MFE only through the number of the
nearest neighbours on the lattice. So, if we would formulate our theory
on the 2-dimensional triangular lattice, we would obtain an effective
dimensionality equal to 3, and for the hexagonal lattice it would be 3/2.
Of course, in the next 1/N correction the dimension of space will enter
in a more serious way, but we expect   even the large N vacuum of QCD
to be less trivial.

On the other hand, none of these features prevent us from hoping for some
meaningful string theory for $D  >  1$
following from the induced gauge theory.
For example,
  the 1D strings \cite{KM1D,Ko1D,1DDS} have an infinite mass spectrum
with an appropriate scaling, and they are a   particular  case of our
model. So, we can hope for the same properties of it for $D > 1$.

Another hope (although a weak one) to attack QCD starting from our
model is the possibility      to reach the physical QCD behaviour
by adjusting      also N to some critical value which is
not separated from $N=\infty$ by a phase transition point,
even though the starting point is  not an asymptotically free theory.
It resembles the hopes related to the strong coupling expansion in
Wilson lattice QCD which died  when  confronted with the computer
simulations. The advantage of the induced gauge theory is that it gives,
at least, most probably starting from the zero approximation
                     the infinite       mass spectrum.

\section{The effective two-matrix model and the MFE based on it}

In this section we propose a new approach to the induced gauge theory
based on its equivalence with an effective 2MM. It is completely
different from the
one presented in our first paper \cite{KM},     which used some
2MM representation     for the IZ integral. The latter was valid for
any 1/n order,
  but the resulting integral equations were difficult to analyse.

This time we will work directly in the large N limit and look from the
very beginning for the homogeneous MF. In this case, as we know,
our effective action is given by (\ref{HomAc}). Note that the    overall
    factor     $(volume)2D$ does not influence the MFE (\ref{MFESum}).
This fact can be used to reduce the problem to an equivalent 2MM with
some modified effective potential.

The 2MM action is a particular case of the action (\ref{Tree})  for two
matrices. In terms of two systems of eigenvalues $\{\phi_k\}$ and
$\{\psi_k\}$, the partition function reads:
\begin{equation}  \matrix{
Z_{2MM} = \int d^N \phi d^N \psi \exp[-N \sum_{k=1}^N (V(\phi_k) +  \cr
V(\psi_k) ) ] \Delta^2(\phi) \Delta^2(\psi) I(\phi,\psi)
}
\label{2MM}
\end{equation}
According to the $Z_2$ symmetry of this model, there should exist a
symmetric saddle point solution $\phi^*_k=\psi^*_k = x_k$.

Comparing the effective action (\ref{2MM}) with the effective action
(\ref{HomAc}) at a homogeneous saddle-point, one finds that the induced
gauge theory is equivalent to the 2MM with (the derivative of) the
effective potential for the latter, which reads
\begin{equation}
V'_{eff}(x_k)= \frac{1}{2D} U'(x_k)+\frac{2D-1}{D N} \sum_{i(\neq k)}
\frac{1}{x_k - x_i}
\label{Veff}
\end{equation}
It is useful to introduce the density of states,  the basic quantity
to be calculated in the large N limit:
\begin{equation}
\rho(x) =      \frac{1}{N } \frac{\partial k(x)}{\partial x}
\label{Dens}
\end{equation}
and the analytical function corresponding to the loop amplitude in 2MM:
\begin{equation}
W(x) = <\tr \frac{1}{x-\Phi}> =
  \int \frac{dy\rho(y)}{x-y}
\label{Loop}
\end{equation}
The effective potential then reads:
\begin{equation}
V'_{eff}(x  )= \frac{1}{2D} U'(x  )+
                                    \frac{2D-1}{D  } Re W(x)
\label{VeffW}
\end{equation}
Hence to find W(x) we have to solve a self-consistent problem: solving
      the 2MM with the effective potential depending on W(x) we find a
self-consistency equation on W(x).     This trick allows us to avoid
the problem of calculation of the IZ integral, since we will now use
a well-elaborated orthogonal polynomial formalism for the 2MM.

Let us review this formalism in the case of an arbitrary potential, using
an elegant approach of the paper \cite{Doug}.

We start from the partition function of 2MM in the form \cite{Mehta}:
\begin{equation}  \matrix{
Z_{2MM} =      \cr
\int d^N x d^N y \Delta(x)\Delta(y)          \cr
\exp \left(N \sum_i(x_i y_i -   V(x_i) -  V(y_i)) \right)
}
\label{TwoMatrix}
\end{equation}
One introduces, according to \cite{Mehta}, the orthogonal polynomials:
\begin{equation}
P_n(x) = x^n + a_{n-1} x^{n-1} + ... +a_0
\label{Polyn}
\end{equation}
obeying the orthogonality condition:
\begin{equation}  \matrix{
<m|n> =  \int d   x d   y       \cr
       \exp \left(\frac{N}{\lambda_0} (x   y   -   V(x  ) -  V(y  ) )
\right)         \cr
  P_m(x)   P_n(y) =
h_n \delta_{mn}
}
\label{Orthog}
\end{equation}
By means of    the Jacobi relations
\begin{equation}    \matrix{
x |n> = \sum_{l \geq -1} x_{n,n+l} |n-l> =
x_n(\hat{z}) |n>          \cr
\frac{\partial}{\partial x} |n> =
   \sum_{l \geq  1} p_{n,n+l} |n-l> = p_n(\hat{z}) |n>
}
\label{Jacobi}
\end{equation}
where $\hat{z} =\exp[-\frac{\partial}{\partial n}]$, one introduces
two operator functions which obey the Heisenberg commutation
relation:
\begin{equation}
[p_n(\hat{z}),x_n(\hat{z})]=1
\label{KdV}
\end{equation}
These relation, can be effectively used in matrix models of 2D gravity
as   a KdV-type approach \cite{Doug,Doug1}.

In the large N limit we introduce the rescaled variables
\begin{equation}  \matrix{
\lambda=\lambda_0\frac{n}{N}     \cr
\omega =-\frac{N}{\lambda_0}\ln(\hat{z}) =
\frac{\partial}{\partial \lambda}
}
\label{ClassV}
\end{equation}
and the KdV-type relation takes the form of the Poisson brackets:
\begin{equation}
\{ p(\lambda,\omega),x(\lambda,\omega)\}=1
\label{ClassKdV}
\end{equation}
where
  $p(\lambda,\omega),x(\lambda,\omega)$  are already ordinary functions.
We could have started from these KdV-type relations, but we will follow
the method of \cite{Doug}.

Considering the matrix element of  $<m|\frac{\partial}{\partial x}|n>$
and integrating by parts inside it, we obtain the equation:
\begin{equation}
p(z)=V'(x(z)) - x\left(\frac{1}{fz}\right)
\label{EquofMot}
\end{equation}
We dropped here the explicit  dependence on the cosmological constant
$\lambda$ and used the transposition operation for $x(z)$
\begin{equation}
x^T(z)=         x(\frac{1}{fz})
\label{Transpoz}
\end{equation}
where $f_n =\frac{h_{n+1}}{h_n}     =    f(\lambda)$

    Eq.(\ref{EquofMot}) is
  in principle  sufficient to define both $x$ and $p$.
Since we know that $p$ contains only positive powers of $z$, we can
compare the coefficients of non-positive powers. This gives sufficient
number of relations to define everything. Particulary simple all this
looks for the polynomial potentials $V(x)$, owing to the    finite sums
in (\ref{Jacobi}). One can, say, easily recover all the results for the
Ising model on random graphs \cite{Mehta,KazI,BuKaz}. In a   forthcoming
paper \cite{DKK}, we will show how effective         equations like
(\ref{ClassKdV}) and (\ref{EquofMot}) are for the complete analysis of
the critical regimes of 2MM.

We can now use eq.(\ref{EquofMot}) for our D-dimensional induced gauge
theory, where we have now to substitute $V(x)$ by $V_{eff}(x)$ from
(\ref{Veff}). We get
\begin{equation}   \matrix{
p(z)=      \cr
\frac{1}{2D} U'(x(z))+
                                      \frac{2D-1}{D  } Re W(x(z))-
                                       x(\frac{1}{fz})
}
\label{EquD}
\end{equation}
To close the equation one has to calculate $W(x)$ in terms of $x(z)$. One
can obtain this relation from an orthogonal polynomial representation of
the eq. (\ref{Loop}). It reads:
\begin{equation}
\frac{\partial W(\lambda,x)}{\partial \lambda} =
\frac{\partial \log  z(\lambda,x)}{\partial x}
\label{Loopxz}
\end{equation}
where $z(x)$ is the function inverse to $x(z)$, to be found from
 (\ref{EquD}).

In the next section we will demonstrate how this  equation  works
for some particular cases.

\section{Particular examples}

Let us demonstrate how the approach worked out in a previous section
works for some particular examples: for usual 2MM, for $D=0$
 (pure 2d-gravity) and for arbitrary D and quadratic potential $U(x)$.

a.Two-Matrix Model: $D=1/2$

In this case $V_{eff}(x) = U(x)$, and we return to the original
 2MM equation (\ref{EquofMot}). A rather complete analysis of it
will be done in \cite{DKK}, so we do not continue on this subject here.

b.Pure Gravity: $D=0$

Now we can retain in (\ref{EquD}) only those terms thath are
singular in $1/D$,       which
immediately gives the well-known equation for the 1-Matrix Model:
\begin{equation}
  2 Re W(x) =      U'(x)
\label{1MM}
\end{equation}
with all familiar consequences for the pure 2D-gravity following from it.

c.Quadratic Bare Potential for any D

In this case we have:
\begin{equation}
                   U'(x)= m^2 x
\label{Quad}
\end{equation}
It is natural to expect the semi-circle law for the distribution of
the eigenvalues here, so we will try the ansatz:
\begin{equation}
W(x)=\frac{1}{2R}\left( x   - \sqrt{x^2-4R} \right)
\label{SCircle}
\end{equation}
where $R(\lambda)$ has to be found.
According to (\ref{Loopxz}), this corresponds to
\begin{equation}
x(z)=\frac{1}{z} + R z
\label{XofZ}
\end{equation}
and
\begin{equation}
ReW(x)=\frac{1}{2R} x
\label{ReW}
\end{equation}
If we insert    all these expressions into (\ref{EquD}) we will find
that $p(\lambda,z)$ should be linear in $z$ and in the normalization
(\ref{ClassV}) it looks like
\begin{equation}
  p(\lambda,z) = \lambda z
\label{PLin}
\end{equation}
Comparing the coefficients of $\frac{1}{z}$ and $z$ in (\ref{EquD}),
we find:
\begin{equation}
 \frac{1}{R}= \frac{m^2(D-1) \pm D \sqrt{m^4 - 2(2D-1)}}{4(2D-1)}
\label{Gross}
\end{equation}
For  $D \geq 1$ and $m^4 \geq m_c^4=2(2D-1)$ we choose, of  corse,
the positiv root. This result coincides with that one obtained in
\cite{Gross} by Migdal's method (see also  \cite{MakLoop}).

This solution is perfectly valid only in the case (which can be called
a strong coupling regime) of real $R$ or for the effective mass M:
\begin{equation}
              M^4 =        m^4 -          2(2D-1) \geq 0
\label{EffM}
\end{equation}
It is clear that the          point $M=0$ corresponds to the critical
behaviour of Bethe tree, when its volume becomes infinite. It is hard
to beleave that some interesting (weak coupling) regime exists beyond
this point.

d) Upside-Down Quadratic Potential for D=1

In this case, as we know (see \cite{KazRev} for     review) the
critical regime corresponds to "upside-down" oscillator   potential,
so we probably have to take the negative root in (\ref{Gross}) and
change $x$ into $i x$ in (\ref{SCircle}), which would give the right
distribution of the eigenvalues for the 1D-matrix model in
the critical regime
describing the 1D-bosonic string theory \cite{KM1D,1DDS}.

\section{Recent developments}

Here we will discuss some alternative approaches to the induced
gauge theory.

a.Migdal's Approach

It is based on an  integral equation found in \cite{Mig}  defining
the large N limit of the Itzykson-Zuber integral. It allowed
some new critical behaviour to be found
for $D>1$ in the model, as well as the
     first attempts in the spectrum calculations to be made.

We will look at this equation
from   a point of view that is a bit different of \cite{Mig},
considering it as a loop equation for the pure 2MM. Then one can
easily restore the whole equation for the D-dimencional theory, by
substituting the bare potential with the effective one, according to
formula (\ref{Veff}).

Consider an obvious identity for the resolvent in 2MM:
\begin{equation}  \matrix{
W(x) =    \cr
\int d^{N^2}\Phi d^{N^2}\Psi     \cr
\exp \left(              - N\tr V(\Phi)-  N \tr V(\Psi)\right)   \cr
\frac{\tr}{N}\frac{1}{x-\partial_\Psi}
\exp      (N \tr\Phi \Psi )
}
\label{Iden}
\end{equation}
Integrating by parts in $\Psi$, we get
\begin{equation}   \matrix{
W(x) =      \cr
\int d^{N^2}\Phi d^{N^2}\Psi
\exp N\tr\left( \Phi \Psi       -   V(\Phi)-  V(\Psi)\right)   \cr
\frac{\tr}{N}\frac{1}{x+\partial_\Psi - V'(\Psi) }1
}
\label{IdeInt}
\end{equation}
The r.h.s. of (\ref{IdeInt}) was calculated in \cite{Mig}  in the
large N limit by means of the Riemann-Cauchi method. The result can
be nicely presented as follows:
\begin{equation}    \matrix{
W(x) =        \cr
- \oint  \frac{dy}{2\pi i} \log \left(x-V'(y)+  W(y) \right)
}
\label{2MMLE}
\end{equation}
where the contour integral encircles the cuts of $W(y)$ but leaves
aside the singularity of the logarithm.

It is             the        loop equation for the 2MM, which was
not known before (see \cite{Alfa} for a similar approach).

It is sufficient to substitute the bare potential with  the effective one
(\ref{Veff}) in order to recover the full D-dimensional equation
of \cite{Mig}.

Using this approach it was found also in \cite{Mig}  that the
 singularity
       of $W(x)$ near the end of the cut should be $x^\alpha$ with
$\alpha = 1  + \frac{1}{\pi} \arccos(\frac{D}{3D-2})$. But the
a   discrepancy with the case D=1/2 (Ising model on random graphs)
where $\alpha=4/3$, is a bit confusing to the author. May be this
solution is aplicable only to $D \geq 1$.

b. Log(log) Critical Behaviour

Another possible solution was advocated in \cite{Bou}.   It was found
that the system for $D \geq 1$ has a critical behaviour of effective
1D-matrix model with the upside-down harmonic potential. The
selfconsistency equation  on the "frequency" leads to a   singularity
of the free energy $\frac{\lambda^2}{\log \log(\lambda)}$
with respect to the cosmological constant.
 This is an interesting
possibility since it would presumably give a well-scaled infinite
mass spectrum, as for the conventional D=1 model. Unfortunately it
can be well justified only in the limit of large D.

The discrepancy between the two approaches is worrysome. It seems that
the model still needs deeper understanding  before we can   view
it as a model of an extended object in physical dimensions.

Some related problems and models can be found also in
   \cite{Dal,Rus}.

Let us note also that a similar model  in two dimensions, but with an
 $F_{\mu\nu}^2$ term, was investigated in \cite{DalKleb}.

\section*{ACKNOWLEDGEMENTS}

I would like to thank D.V.Boulatov, J.-M.Daul, and I.K.Kostov for
interesting discussions, and the organizers of "LATTICE-92" for
their kind hospitality.

%section*{REFERENCES}


\begin{thebibliography}{31}

\bibitem{KM} V.A.Kazakov and A.A.Migdal, {\it Induced QCD at large N},
preprint LPTENS-92/15, PUPT-1322 (June 1992); Nucl.Phys.B, to be
published.

\bibitem{Sem}   I.I.Kogan, G.W.Semenoff and N.Weiss, {\it Induced QCD and
hidden local $Z_N$ symmetry}, UBC preprint UBCTP-92-022 (June 1992);\\
I.I.Kogan, G.W.Semenoff, A.Morozov and N.Weis, {\em Area law and continuum
limit in induced QCD}, preprint UBCTP-92/22 (July 1992);\\
I.I.Kogan, G.W.Semenoff, A.Morozov and N.Weis, {\em Continuum Limits of
limits of  "induced QCD"}, preprint UBCTP-92/27 (August 1992);\\

S.Khokhlachov and Yu.Makeenko, {\it The problem of
large-N phase transition in Kazakov-Migdal model of induced QCD}, ITEP
preprint ITEP-YM-5-92 (July 1992);\\

\bibitem{BouSym} D.V.Boulatov, {\it Local symmetry in the Kazakov-Migdal
gauge model}, preprint NBI-HE-92-62 (September 1992);

\bibitem{MaKh}
S.Khokhlachov and Yu.Makeenko,  ITEP preprint ITEP-YM-7-92 (July 1992);\\

\bibitem{MigFerm} A.A.Migdal,
{\it Mixed models of induced QCD} preprint LPTENS-92/23 (August 1992)\\
  Princeton preprint PUPT-1343 (September 1992);

\bibitem{Kpot}  V.A.Kazakov, Nucl.Phys.B (Suppl.)                (1987).

\bibitem{Bou}  D.V.Boulatov,  preprint NBI-HE-92-78  (November 1992);

\bibitem{Mig}  A.A.Migdal, {\it Exact solution of induced lattice gauge
theory at large N}, Princeton preprint PUPT-1323 (June 1992);\\
preprint LPTENS-92/22 (August 1992);\\
 {\it 1/N
expansion and particle spectrum in induced QCD}, Princeton preprint
PUPT-1332 (July 1992); {\it Phase transitions in induced QCD}, ENS
preprint LPTENS-92/22 (August 1992);

\bibitem{Gross} D.Gross, {\it Some remarks about induced QCD}, Princeton
preprint PUPT-1335 (August 1992);\\

\bibitem{MakLoop}
Yu.Makeenko, {\it Large $N$ reduction, Master Field and Loop Equations
in Kazakov-Migdal model}, preprint ITEP-YM-6-92 (August 1992);

\bibitem{Casel} M.Caselle, A.D'Adda and S.Panzeri, preprint DFTT 38/92
(July 1992);

\bibitem{MonteC} A.Glocksch and Yue Shen, preprint BNL, (July   1992);\\
S.Aoki, A.Gocksch and Y.Shen, preprint UTHEP-242, August (1992).

\bibitem{KaQCD} V.A.Kazakov, Phys.Lett.128B, 316 (1986).

\bibitem{KoQCD} I.K.Kostov,  Nucl.Phys.B265 [FS15], 223 (1986).

\bibitem{IZ}    C.Itzykson and J.-B.Zuber, {\sl J.Math.Phys.} {\bf 21}
(1980) 411;

\bibitem{Doug} M.R.Douglas, proceedings of Cargese Workshop 1990,
in "Random Surfaces and Quantum Gravity", ed. by O.Alvarez, et al.
(1991).

\bibitem{Mehta} M.L.Mehta, Comm.Math.Phys. 79, 327  (1981).

\bibitem{Doug1} M.R.Douglas, Phys.Lett.B238, 176 (1990).

\bibitem{KazI} V.A.Kazakov, Phys.Lett. 119A, 140 (1985).

\bibitem{BuKaz} D.V.Boulatov and V.A.Kazakov, Phys.Lett. 186B, 379(1987).

\bibitem{DKK} J.-M.Daul, V.A.Kazakov and I.K.Kostov, to appear.


\bibitem{KM1D}   V.A.Kazakov and A.A.Migdal, Nucl. Phys. {\bf B311}
(1988) 171;

\bibitem{Ko1D} I.K.Kostov, Phys.Lett.B215, 499 (1988).

\bibitem{1DDS}     E.Brezin, V.A.Kazakov,    Al.B.Zamolodchikov,
Nucl. Phys. {\bf B338} (1990) 673;\\
D.Gross and N.Milkovic, Phys. Lett. {\bf B238} (1990) 217;\\
P.Ginsparg and J.Zinn-Zustin, Phys. Lett. {\bf B240} (1990) 333;
G.Parisi, Phys.Lett.

\bibitem{KazRev} V.A.Kazakov,
             proceedings of Cargese Workshop-1990,
in "Random Surfaces and Quantum Gravity", ed. by O.Alvarez, et al.(1991).

\bibitem{Alfa} J.Alfaro, preprint CERN-TH-6531/92, (July, 1992).

\bibitem{Rus} B.Rusakov, {\em From hermitean matrix model
to lattice gauge theory}, preprint TAUP 1996-92 (September, 1992)

\bibitem{Dal} S.Dalley, preprint PUPT-1310, (March 1992).

\bibitem{Dal} S.Dalley and I.Klebanov,  preprint PUPT-1333  (July  1992).

\bibitem{DalKleb} S.Dalley and I.Klebanov,  preprint PUPT-1342
 (September 1992).

\end{thebibliography}
\end{document}